\documentclass[preprint,prd,showpacs]{revtex4} 
 
\begin{document} 
\input{epsf}

\title{Negative Energy in Superposition and Entangled States}

\author{ L.H. Ford}
 \email[Email: ]{ford@cosmos.phy.tufts.edu} 
 \affiliation{Institute of Cosmology  \\
Department of Physics and Astronomy\\ 
         Tufts University, Medford, MA 02155}
\author{Thomas A. Roman}
  \email[Email: ]{roman@ccsu.edu}
  \affiliation{Department of Mathematical Sciences \\
 Central Connecticut State University \\  
New Britain, CT 06050}  

\begin{abstract}
We examine the maximum negative energy density which can be attained
in various quantum states of a massless scalar field. We consider
states in which either one or two modes are excited, and show that the
energy density can be given in terms of a small number of parameters.
We calculate these parameters for several examples of superposition states
for one mode, and entangled states for two modes, and find the maximum 
magnitude of the negative energy density in these states. We consider 
several states which have been, or potentially will be, generated in 
quantum optics experiments.
\end{abstract}

\maketitle

\section{Introduction}
\label{sec:intro}

It has been proven, beginning in the early 1960's \cite{EGJ}, that there 
always 
exist states with negative energy density in quantum field theory. 
Some specific examples 
include the Casimir effect \cite{C} and squeezed states \cite{SY}, 
both of which have 
been experimentally realized. (Although the energy density itself is far too 
small to be directly measured.) Negative energy is also required for black hole
evaporation, and hence for the consistency of the laws of black hole physics with those 
of thermodynamics. 
On the other hand, unrestricted amounts of negative energy 
could produce bizarre effects, for example, violations of the 
second law of thermodynamics \cite{F78,F91}. However, the same laws of 
quantum field theory which allow the existence of negative energy also 
appear to severely restrict its magnitude and duration in such a way as to prevent 
gross large-scale effects. These bounds are known as quantum inequalities, and 
quite a large body of work now exists on the subject. For some recent 
reviews of quantum inequalities, see Refs.~\cite{LF100,TRMGM,CJF}. Quantum inequality 
bounds have been proven, for example, for the minimally coupled scalar, electromagnetic, 
and Dirac fields. 
It should be pointed out that the potential macroscopic problems arise not because 
of the existence of negative energy per se, but from the arbitrary separation of negative 
and positive energy. It is this behavior which the quantum inequalities prohibit. 
Many possible configurations of separated negative and positive energy can easily be ruled 
out, and known permitted examples involve the subtle intertwining of the two \cite{BFR}. 
Whether the currently known examples are representative of the general case is unknown. 
Hence, the study of further examples could prove useful. 
 
Since the negative energy densities in these states are too small to be directly measurable, 
experiments in quantum optics may offer the best possibilities for indirect detection 
of negative energy. (However, see also Refs.~\cite{FGO,DO}.) 
A first link between quantum optics and the work on quantum inequalities 
has been forged in a recent paper by Marecki \cite{PM}. 
For squeezed states, he proved quantum inequality-type bounds 
on the magnitude and duration of the squeezing. 

Quantum optics has seen enormous experimental and theoretical advances in the 
last twenty years. This marriage of optics with quantum field theory has 
resulted 
in experiments which were formerly purely ``gedanken'' becoming those which 
are now routinely performed in the laboratory. Highly non-classical states, 
such as Schr{\"o}dinger ``cat states'' and squeezed states, 
have been produced and play a part in everything from 
quantum computers to noise reduction in laser interferometer gravitational 
wave detectors. The ``cat states'' of the electromagnetic field are superpositions of
coherent states and have been created experimentally~\cite{Auffeves,Pathak}. The experiments
which have been done so far have produced mesoscopic superpositions, in which the mean photon
number is of order $10$. This is somewhat short of a true Schr{\"o}dinger cat state, which
would be a superposition of two or more classical configurations, that is, coherent states
with very large occupation numbers.  More recently, there have been proposals for methods of creating
superpositions of squeezed vacuum states~\cite{Zhang}. 

 An interesting 
question arises: can one start with two quantum states which do not involve negative energy 
and by superposing them obtain negative energy? The answer is yes; the classic standard 
example being the vacuum $+$ two-particle state (for a nice discussion see Ref.~\cite{MP-thesis}). 
Is this true for the superposition of other states as 
well? More generally, what effects does 
superposition have on negative energy? Could one also go the other way, i.e., 
start with two states involving negative energy and by superposing them 
diminish or eradicate the negative energy? In this paper, we will address such questions for
several classes of states. In Sect.~\ref{sec:rho}, we develop some formalism for parameterizing
the maximum magnitude of negative energy that can occur for states of a minimally coupled scalar field 
in Minkowski spacetime with either one or two modes excited. We give several examples of superpositions
for a single mode in Sect.~\ref{sec:super}, including superpositions of two coherent states, 
two squeezed vacuum states, and a coherent state with a squeezed vacuum state. In Sect.~\ref{sec:entangled},
we move to the two-mode case. This allows us to consider examples of entangled states involving either
squeezed vacua or coherent states for the two modes.
 A summary of our conclusions is presented in Section~\ref{sec:Sum}.

\section{Energy Density with One or Two Modes are Excited}
\label{sec:rho}

In this paper, we will consider a massless scalar field in 
flat spacetime, for which the stress tensor operator is
\begin{equation}
T_{\mu\nu} = \varphi_{,\mu}\,\varphi_{,\nu}\, - \frac{1}{2} g_{\mu\nu}\,
              \varphi_{,\sigma} \varphi^{,\sigma}\,.
\end{equation} 
The normal-ordered energy density operator is 
\begin{equation}
:T_{00}: = \frac{1}{2} [:{\dot \varphi}^2: + :(\nabla \varphi)^2:] \,,
\end{equation}
where
\begin{equation}
\varphi = {\sum_k} ({a_k}{f_k} + {a_k}^\dagger{f_k}^\ast) \,,
\end{equation}
with the ${f_k}({\bf x},t)$ being the mode functions.   

\subsection{Two Modes Excited}
\label{eq:two}

We wish to consider the case where all modes except for two are in the vacuum
state. For the first mode, let $f_1$, $a$, and $a^\dagger$ be the mode 
function, annihilation operator, and creation operator, respectively, and let 
 $f_2$, $b$, and $b^\dagger$ be the corresponding quantities for the second
mode. The expectation value of the energy density in an arbitrary quantum
state can be expressed as
\begin{eqnarray}
\rho &=& \langle :T_{00}: \rangle = 
{\rm Re} \biggl\{ \langle a^\dagger a \rangle\,
(|{\dot f_1}|^2 +|{\nabla f_1}|^2) +  \langle a^2 \rangle\,
[{\dot f_1}^2 +({\nabla f_1})^2] +
\langle b^\dagger b \rangle\,
(|{\dot f_2}|^2 +|{\nabla f_2}|^2)   \nonumber \\
&+&  \langle b^2 \rangle\,
[{\dot f_2}^2 +({\nabla f_2})^2]  
 + 2 \langle a^\dagger b \rangle\, ({\dot f_1}^* {\dot f_2} +
{\nabla f_1}^*  \cdot {\nabla f_2}) + 
2 \langle a b \rangle\, ({\dot f_1} {\dot f_2} +
{\nabla f_1} \cdot {\nabla f_2}) \biggr\} \,.  \label{eq:rho}
\end{eqnarray}
Let 
\begin{equation}
n_1 = \langle a^\dagger a \rangle, \quad
n_2= \langle b^\dagger b \rangle, \quad
R_1\, {\rm e}^{i\gamma_1} =  \langle a^2 \rangle, \quad
R_2\, {\rm e}^{i\gamma_2} = \langle b^2 \rangle, \quad
R_3\, {\rm e}^{i\gamma_3} = \langle a^\dagger b \rangle, \quad
R_4\, {\rm e}^{i\gamma_4} = \langle a b \rangle\,.
\end{equation}
All of the information needed to give the two-mode energy density,
 Eq.~(\ref{eq:rho}), at a given quantum state is encoded in the above set
of six amplitudes and four phases. 

In the case of a
traveling waves, we may take the mode functions to be
\begin{equation}
f_j = {i \over \sqrt{2\omega_j V}} 
{\rm e}^{i(\mathbf{k}_j \cdot \mathbf{x} - \omega_j t)} \,,
\end{equation}
where $\omega_j = |\mathbf{k}_j|$, for $j=1,2$ and $V$ is a normalization 
volume. In this case, the mean energy density may be expressed as
\begin{eqnarray}
\rho &=& \frac{1}{V}\,\biggl\{ n_1\, \omega_1 + n_2\, \omega_2
+ R_1\, \omega_1\, \cos[2(\mathbf{k}_1 \cdot \mathbf{x} - \omega_1 t)+\gamma_1]
+ R_2\, \omega_2\, \cos[2(\mathbf{k}_2 \cdot \mathbf{x} - \omega_2 t)+\gamma_2]
\nonumber \\ 
   &+& R_3\, \sqrt{\omega_1 \omega_2}\, 
(1+ \mathbf{\hat{k}}_1 \cdot \mathbf{\hat{k}}_2)\,
\cos[(\mathbf{k}_2-\mathbf{k}_1) \cdot \mathbf{x} -(\omega_2-\omega_1)t 
+\gamma_3]    \nonumber \\
  &+& R_4\, \sqrt{\omega_1 \omega_2}\, 
(1+ \mathbf{\hat{k}}_1 \cdot \mathbf{\hat{k}}_2)\,
\cos[(\mathbf{k}_2+\mathbf{k}_1) \cdot \mathbf{x} -(\omega_2+\omega_1)t 
+\gamma_4] \biggr\} \,. \label{eq:2mode_rho}
\end{eqnarray}
We will also consider the case of a standing wave which depends upon only 
one space coordinate, in which case the mode functions can be taken to be
\begin{equation}
f_j = {1 \over \sqrt{\omega_j V}} \, \sin(\omega_jx)\, 
{\rm e}^{ -i \omega_j t} \,.
\end{equation}
The energy density now becomes
\begin{eqnarray}
\rho &=& \frac{1}{V}\,\biggl\{ n_1\, \omega_1 + n_2\, \omega_2
+ R_1\, \omega_1\, \cos(2 \omega_1 x)\, \cos(2\omega_1 t -\gamma_1)
+ R_2\, \omega_2\, \cos(2 \omega_2 x)\, \cos(2\omega_2 t -\gamma_1)
\nonumber \\ 
   &+& 2R_3\, \sqrt{\omega_1 \omega_2}\, 
\cos[(\omega_2-\omega_1)x] \, \cos[(\omega_2-\omega_1)t -\gamma_3] 
 \nonumber \\ 
&+& 2R_4\,  \sqrt{\omega_1 \omega_2}\, 
\cos[(\omega_1+\omega_2)x] \, \cos[(\omega_1+\omega_2)t -\gamma_4]  \biggr\}
 \,. \label{eq:2mode_rho_stand}
\end{eqnarray}

\subsection{One Mode Excited}
\label{eq:one}

A useful special case is when only one mode is excited. In this case, we may
set $n_1 = n$, $R_1=R$, $\gamma_1=\gamma$, and $R_2=R_3=R_4=\gamma_2=
\gamma_3=\gamma_4=0$. In this case, we need only the three real numbers
$n$, $R$, and $\gamma$ to determine the energy density in a given state.
For the case of a traveling wave, we have
\begin{equation}
\rho = \frac{\omega}{V}\, \left\{  n + R\, 
\cos([2 (\mathbf{k} \cdot \mathbf{x}- \omega t) + \gamma] \right\}\, 
  \label{eq:rho1}
\end{equation}
We can see from Eq.~(\ref{eq:rho1}) that the minimum value of $\rho$ is
\begin{equation}
\rho_{min} = - \frac{\omega}{V}\, ( R - n ) \,,
                                             \label{eq:rho_min}
\end{equation}
and hence we can have negative energy density only if $R >  n $.
In the case of a standing wave, Eq.~(\ref{eq:2mode_rho_stand}) becomes
\begin{equation}
\rho = \frac{\omega}{V}\, \left[  n  + R\, \cos(2 \omega x) \,
\cos(2\omega t - \gamma) \right]\,.   \label{eq:rho2}
\end{equation}
Again, the minimum value of $\rho$ is given by Eq.~(\ref{eq:rho_min}).

\section{Superpositions for One Mode}
\label{sec:super}

In this section, we examine some explicit examples of superpositions involving
a single mode. In each case, we need only calculate the quantity $R-n$
to determine the maximum magnitude of the negative energy.

\subsection{Superposition of Two Coherent States}
\label{sec:SCS}

First we consider a superposition of coherent states. 
Coherent states are eigenstates of the 
annihilation 
operator, that is
\begin{equation}
a| \alpha \rangle = \alpha | \alpha \rangle \,.
\end{equation} 
Let 
\begin{equation}
\psi \rangle = N [|\alpha \rangle + \eta |\beta \rangle] \,,
\label{eq:SCS}
\end{equation}
where $|\alpha \rangle$ and $|\beta \rangle$ are two different coherent 
states for 
the same mode, $\eta$ is a complex number, and $N$ is a normalization 
factor 
(see, for example, Sec. 7.6 of Ref.~\cite{GK}). We also assume that the 
states are normalized so that 
$\langle \alpha |\alpha \rangle =\langle \beta |\beta \rangle = 1$. 
As a result we 
have that 
\begin{equation}
\langle \psi | \psi \rangle = 1 = 
N^2 [1 + {|\eta|}^2 + \eta \langle \alpha| \beta \rangle 
+ \eta^\ast \langle \beta| \alpha \rangle] \,.
\end{equation}
The coherent states are not orthonormal; their overlap integral is given by 
(see for example, Eq.(3.6.24) of Ref.~\cite{BR}):
\begin{equation}
\langle \alpha| \beta \rangle = e^{-\frac{1}{2}({|\alpha|}^2 + 
{|\beta|}^2 - 2 \alpha^\ast \beta)} \,.
\end{equation}
Therefore the square of the normalization factor is 
\begin{equation}
N^2 = {[1+ {|\eta|}^2 + 2 {\rm e}^{-\frac{1}{2}({|\alpha|}^2 + 
{|\beta|}^2)} \, {\rm Re}(\eta {\rm e}^{ \alpha^\ast \beta})]}^{-1} \,.
\end{equation}

The mean number of particles is found to be
\begin{equation}
n = N^2 \left[ |\alpha|^2 + |\eta \beta|^2 +
 2 {\rm e}^{-\frac{1}{2}({|\alpha|}^2 + {|\beta|}^2)} \, 
{\rm Re}(\eta \alpha^\ast \beta {\rm e}^{ \alpha^\ast \beta})\right]\,,
\end{equation}
and 
\begin{equation}
\langle a^2 \rangle =  N^2 \left[ \alpha^2 + |\eta|^2 \beta^2 +
{\rm e}^{-\frac{1}{2}({|\alpha|}^2 + {|\beta|}^2)} \,
(\eta \beta^2 {\rm e}^{ \alpha^\ast \beta}
+\eta^\ast \alpha^2 {\rm e}^{ \alpha \beta^\ast}) \right]\,.
\end{equation}
Let
\begin{equation}
\alpha = |\alpha| {\rm e}^{i\delta_1}\,, \quad 
\beta = |\beta| {\rm e}^{i\delta_2}\,, \quad {\rm and} \quad
\eta = |\eta|  {\rm e}^{i\delta}\,.
\end{equation}
Then the quantities $n$, $R$, and $\gamma$ are functions of six real 
parameters, the magnitudes and phases of $\alpha$, $\beta$, and $\eta$.
However, one finds that only $\gamma$ depends upon all six. 
The magnitudes $n$ and $R$
depend only upon the difference $\delta_2-\delta_1$, and are hence functions 
of five parameters. We are primarily interested in the quantity $R-n$, which
measures the maximum magnitude of the negative energy density. Hence set
$\delta_1 =0$ and write
\begin{equation}
F(|\alpha|, |\beta|, |\eta|, \delta_2, \delta) = R -n \,, \label{eq:F}
\end{equation}
and let $G$ be a five-dimensional vector given by
\begin{equation}
G = \left(\frac{\partial F}{\partial |\alpha|},
\frac{\partial F}{\partial |\beta|},
\frac{\partial F}{\partial |\eta|},
\frac{\partial F}{\partial \delta_2},
\frac{\partial F}{\partial \delta} \right)\,.  \label{eq:G}
\end{equation}

One may use Eqs.~(\ref{eq:F}) and (\ref{eq:G}) as the basis of 
a numerical algorithm to search for points of maximum $F$ and hence maximally
 negative energy density. Start at a random point in the five-dimensional
parameter space, and compute $F$ and $G$. If $F>0$, then this choice of
parameters is a quantum state with negative energy density. The components
of $G$ indicate the direction in which $F$ is increasing most rapidly. One 
then moves along this direction until a local maximum of $F$ is located.
A preliminary, non-exhaustive, search located two such local maxima, at
$(|\alpha|, |\beta|, |\eta|, \delta_2, \delta) \approx (0.8,0.8,1,3.14,0)$
and at $(|\alpha|, |\beta|, |\eta|, \delta_2, \delta) \approx (0,1.61,1,0,0)$.
(One can trivially generate a third maximum by interchange of $\alpha$ and
$\beta$ in the latter case.) The first example corresponds to $\alpha = -\beta
=0.8$ and the second to a superposition of a coherent state and the vacuum.
Interestingly, the maximum magnitude of the  negative energy density is about
the same in both examples, with $F = R -n \approx 0.278$, and hence 
$\rho_{min} \approx - 0.278\, \omega/V$.   We do not have an 
explanation  as to why these two choices give the same value of $R-n$. The
mean particle number is $n \approx 0.36$ in the first example and 
$n \approx 1.0$ in the second. This example illustrates that a superposition 
of two coherent states can produce negative energy density, and the maximum
negative energy density arises for mean particle number of order one.

\subsection{Superposed Squeezed Vacuum States}
\label{sec:ESVS}

\subsubsection{A Single-Mode Squeezed Vacuum State}
\label{sec:single_sq}

We begin with a review of the features of the 
expectation value of the energy density in a single squeezed vacuum state. 
Our state is given by:
\begin{equation}
| \psi \rangle = | \xi \rangle \,\,\,\,\,\,,\,{\rm with} \,\, \xi = r \, e^{i \delta} \,,
\end{equation} 
where $r$ is the squeeze parameter and $\delta$ is a phase parameter.
The squeeze operator $S(\xi)$ is given by
\begin{equation}
S(\xi) = e^{\frac{1}{2} [\xi a^2 - \xi^\ast {(a^\dagger)}^2]} \,.
\end{equation}
This operator is unitary since 
\begin{equation}
S^\dagger (\xi)  = S(-\xi)=S^{-1}(\xi)\,.
\end{equation}
The single-mode squeezed state $|\xi \rangle$ is produced by the squeeze operator acting 
on the vacuum state
\begin{equation}
|\xi \rangle = S(\xi) |0 \rangle \,.
\end{equation}
The state $|\xi \rangle$ can be written, after some work 
(see Eq. (3.7.5) of Ref.~\cite{BR}), in terms of the even Fock states as 
\begin{equation}
|\xi \rangle = \sqrt{{\rm sech}r} \, \sum_{n=0}^\infty \, 
\frac{\sqrt{(2n)!}}{n!} 
{\biggl[-\frac{1}{2} e^{i \delta} {\rm tanh}r \biggr]}^n \,|2n \rangle \,.
\label{eq:xi}
\end{equation}
We also have that
\begin{eqnarray}
S^\dagger (\xi) a S(\xi) &=& a \,{\rm cosh}r - 
a^\dagger e^{i \delta} {\rm sinh}r \,, \nonumber \\
S^\dagger (\xi) a^\dagger S(\xi) &=& a^\dagger {\rm cosh}r - 
a\, e^{-i \delta} {\rm sinh}r \,.
\label{eq:SdagaS}
\end{eqnarray}
In the state $|\xi \rangle$ we have the expectation values
\begin{eqnarray}
n &=& \langle a^\dagger a \rangle = \langle 0| S^\dagger (\xi) a^\dagger a 
S(\xi) |0 \rangle =
\langle 0| S^\dagger (\xi) a^\dagger S(\xi) S^\dagger (\xi) a S(\xi)
 |0 \rangle 
= {\rm sinh}^2 r \,, \nonumber  \\
\langle a^2 \rangle &=& \langle 0| S^\dagger (\xi) a^2 S(\xi) |0 \rangle =
\langle 0| S^\dagger (\xi) a S(\xi) S^\dagger (\xi) a S(\xi) |0 \rangle 
= -e^{i \delta}{\rm sinh}r \, {\rm cosh}r \,, \label{eq:diag_sq}
\end{eqnarray}
where we have made use of Eqs.~(\ref{eq:SdagaS}). Thus $R= \sinh r\;\cosh r$
and 
\begin{equation}
R-n = \sinh r (\cosh r - \sinh r)  \label{eq:Rnsqvac}
\end{equation}
attains its maximum value of $0.5$ as $r \rightarrow \infty$.

\subsubsection{Superposition of Squeezed Vacuum States}

We now calculate the energy density in a superposition of two single-mode 
squeezed vacuum states of the form 
\begin{equation}
|\psi \rangle = N [|\xi \rangle + \eta |-\xi \rangle] \,,
\label{eq:ESVS}
\end{equation} 
where, for simplicity, we will choose
\begin{equation}
\xi = r, \,\,\,\,\,\,\,\,\,\,\delta = 0 \,,
\end{equation}
and set
\begin{equation}
\eta = |\eta|\, {\rm e}^{i \theta}\,.
\end{equation}
In this state we have
\begin{eqnarray}
n = \langle a^\dagger a \rangle &=& 
N^2 [\langle \xi | a^\dagger a| \xi \rangle 
+ {|\eta|}^2 \,  \langle -\xi | a^\dagger a| -\xi \rangle 
+ \eta \, \langle \xi | a^\dagger a| -\xi \rangle 
+ \eta^\ast \, \langle -\xi | a^\dagger a| \xi \rangle] \nonumber \\
&=& N^2 \biggl[{\rm sinh}^2 r (1+{|\eta|}^2) 
- 2 |\eta|\, \cos \theta  \, \frac{{\rm sech}r \,
{\rm tanh}^2 r}{{(1+{\rm tanh}^2 r)}^{3/2}} \biggr] \,, \label{eq:n_2sq}
\end{eqnarray}
where we have made use of Eq.~(\ref{eq:ada}) in the Appendix. 
A similar calculation,  using Eq.~(\ref{eq:a2}),   yields
\begin{equation}
\langle a^2 \rangle =  
N^2 \biggl[({|\eta|}^2 - 1)\, {\rm sinh}r \,{\cosh}r
+ 2\, i |\eta|\, \sin \theta \, \frac{{\rm sech}r \,
{\rm tanh}r}{{(1+{\rm tanh}^2 r)}^{3/2}} \biggr] \,, 
\end{equation}
and
\begin{equation}
R = |\langle a^2 \rangle| =  
N^2 \biggl[({|\eta|}^2 - 1)^2\, {\rm sinh}^2r \,{\cosh}^2r
+ 4 |\eta|^2\, \sin^2 \theta \, \frac{{\rm sech^2}r \,
{\rm tanh^2}r}{{(1+{\rm tanh}^2 r)}^{3}} \biggr]^\frac{1}{2} \,. 
\label{eq:R_2sq}
\end{equation}
The normalization of our state is given by
\begin{equation}
{\langle \psi | \psi \rangle} = 
1 = N^2 \,[1+{|\eta|}^2 + \eta \langle\xi|-\xi\rangle 
+\eta^\ast \langle -\xi|\xi\rangle] \,.
\end{equation}
The square of the normalization factor is then 
\begin{equation}
N^2 = {\biggl[1+{|\eta|}^2 + 2 |\eta|\, \cos \theta  \,
\sqrt{{\rm sech}(2r)}\, \biggr]}^{-1} \,,
\label{eq:Nsquared}
\end{equation}
where we have used Eq.~(\ref{eq:sech}) of the Appendix.

\begin{figure} 
\begin{center} 
\leavevmode\epsfysize=8cm\epsffile{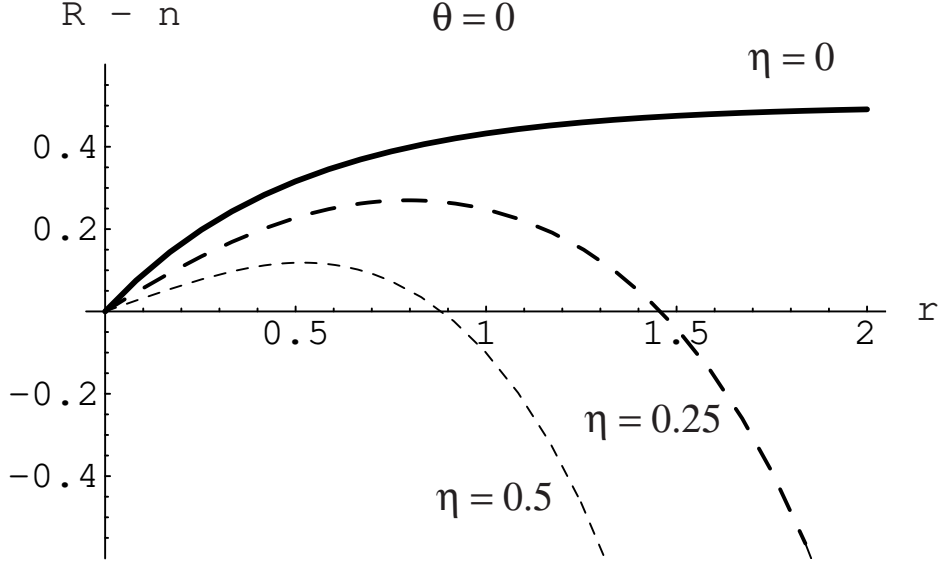} 
\end{center} 
\caption{The quantity $R-n$ for two superposed squeezed vacua, 
Eq.~(\ref{eq:ESVS}),
is plotted for the case $\theta=0$ for various values of $\eta$. The
case $\eta=0$ is the single squeezed vacuum, and gives more negative energy
density than do any of the superpositions. For non-zero $\eta$, there
is a maximum value for $R-n$ at a finite value of $r$.   } 
\label{fig:2sq} 
\end{figure} 

From Eqs.~(\ref{eq:n_2sq})  and (\ref{eq:R_2sq}), 
we can compute the quantity $R-n$, which gives the 
maximum magnitude of the negative energy density, as function of $\theta$
and $r$. For fixed $\theta$, one typically finds that $R-n$ attains a 
maximum value for some value of $r$, usually of order one. A typical
case of $\theta=0$ is illustrated in Fig.~\ref{fig:2sq}. 
The case $\eta=0$ is just 
the single squeezed vacuum state discussed in Sect.~\ref{sec:single_sq}  . 
This case gives 
the maximum negative energy density, $R-n = 0.5$, for large $r$. All other 
values of $\eta$, corresponding to superposed squeezed vacua, give somewhat 
smaller amounts of negative energy density, and attain their maximum negative
energy density at finite values of $r$.

\subsection{Superposition of Coherent and Squeezed Vacuum States}
\label{sec:CSV}

In this subsection, we consider states of the form
\begin{equation}
|\psi \rangle = N [|\xi \rangle + \eta |\alpha \rangle]\,, \label{eq:csv}
\end{equation}
where $|\xi \rangle$ is a squeezed vacuum state, and $ |\alpha \rangle$ is
a coherent state. We may use Eq.~(\ref{eq:in_prod}) to find
\begin{equation}
N^2 = \left\{1+|\eta|^2 +2 \sqrt{{\rm sech}\, r}\, 
{\rm e}^{-\frac{1}{2} |\alpha|^2}
\; {\rm Re}\left[\eta \exp\left(-\frac{1}{2}\, {\rm e}^{-i \delta}\,
\alpha^2\, \tanh r\right) \right]\right\}^{-1} \,.
\end{equation}
Similarly, we find
\begin{eqnarray}
n &=& \langle a^\dagger a \rangle = N^2\, \biggl\{ \sinh^2 r  +|\eta \alpha|^2
    \nonumber \\  
&-&2  {\rm e}^{-\frac{1}{2} |\alpha|^2} \sqrt{{\rm sech}\, r}\,\tanh r\;
{\rm Re}\left[\eta \, {\rm e}^{-i \delta}\, \alpha^2\, 
\exp \left(-\frac{1}{2}\, {\rm e}^{-i \delta}\,
\alpha^2\, \tanh r \right) \right]\biggr\}
\end{eqnarray}
and 
\begin{eqnarray}
\langle a^2 \rangle &=& N^2\, \biggl\{- \sinh r \cosh r \, {\rm e}^{i \delta}  
+|\eta|^2  \alpha^2  + \eta \, \alpha^2\,\sqrt{{\rm sech}\, r}\,
{\rm e}^{-\frac{1}{2} |\alpha|^2} \; 
\exp\left(-\frac{1}{2}\, {\rm e}^{-i \delta}\,
\alpha^2\, \tanh r \right) \nonumber \\
&+& \eta^* \,\sqrt{{\rm sech}\, r}\, {\rm e}^{-\frac{1}{2}  |\alpha|^2}
[(\alpha^*)^2 \, {\rm e}^{i \delta}\, \tanh r -1]\, 
{\rm e}^{i \delta}\, \tanh r \;
\exp[-\frac{1}{2}\, {\rm e}^{i \delta}\,(\alpha^*)^2\, \tanh r ] \biggr\}\,,
\end{eqnarray}
using Eqs.~(\ref{eq:ada2}) and (\ref{eq:ad2}).

\begin{figure} 
\begin{center} 
\leavevmode\epsfysize=8cm\epsffile{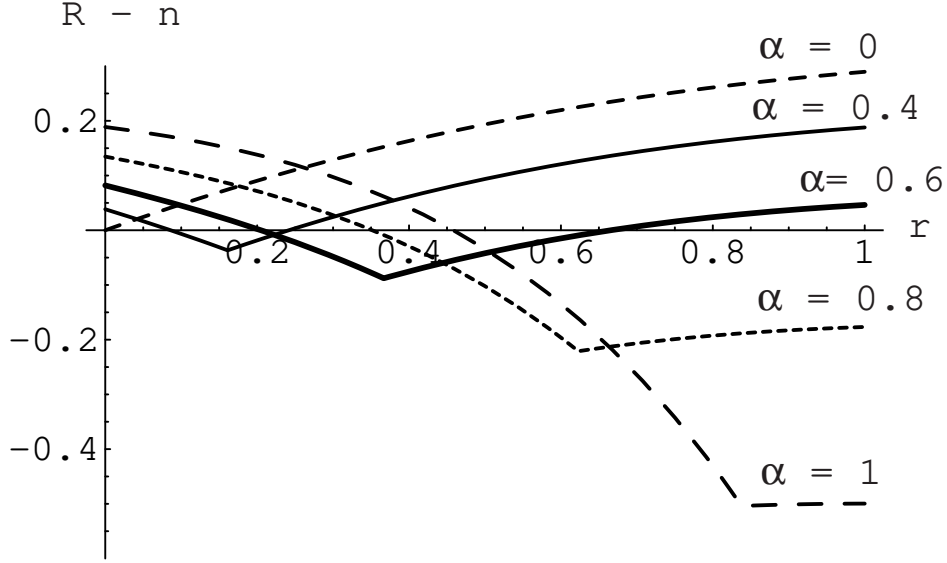} 
\end{center} 
\caption{Here $R-n$ is plotted for the superposition of a coherent 
and a squeezed
vacuum state, Eq.~(\ref{eq:csv}) with $\eta = 1$, 
as a function of $r$ for various values of 
$\alpha$. Note that for $\alpha \not= 0$, $R-n$ can be positive, 
corresponding to negative energy, for both smaller and larger values of 
$r$, but has an intermediate region where there is no negative energy.   } 
\label{fig:scs} 
\end{figure}

Let us consider the case where $\delta=0$, $\eta 
= 1$, and $\alpha$ is real. In this case, $R-n$ is plotted in 
Fig.~\ref{fig:scs}
for various values of $\alpha$. Here the maximum
negative energy density, $R-n \approx 0.23$ is attained for large $r$.
Note that this state has less negative energy than does the squeezed 
vacuum by itself. [See Eq.~(\ref{eq:Rnsqvac}).] 
For $\alpha$ non-zero, we find that $R-n$ initially
decreases, reaches a minimum value, and then increases again. For the case
$\alpha =0.6$, for example, there is negative energy for $r < 0.2$ and 
again for $r > 0.65$, but not for intermediate values of $r$. Note that
$r=0$ is a superposition of the vacuum and a coherent state, a special 
case of the state treated in Sect.~\ref{sec:SCS}.

\begin{figure} 
\begin{center} 
\leavevmode\epsfysize=8cm\epsffile{ 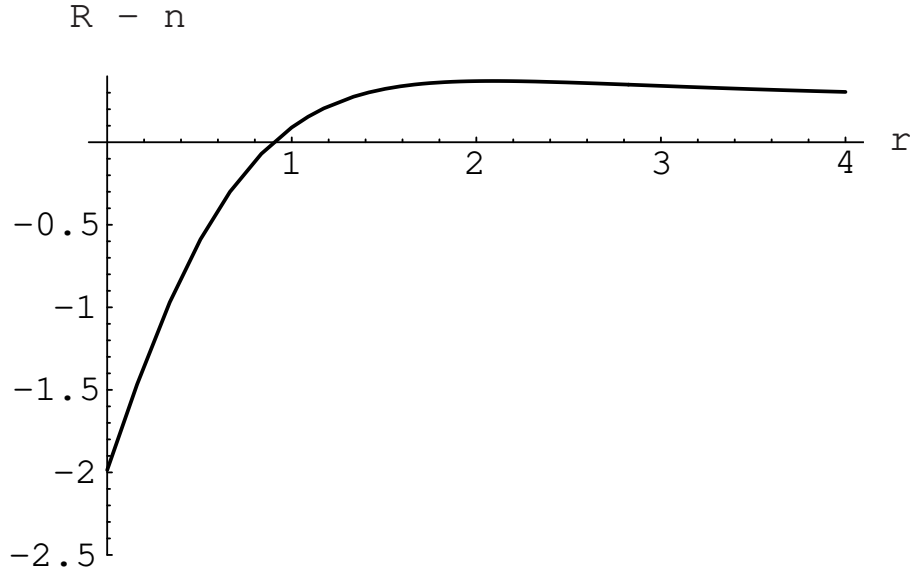} 
\end{center} 
\caption{Here $R-n$ is plotted as a function of $r$ for $\alpha=0$ and 
$\eta=-1$, a superposition of the vacuum and a squeezed vacuum. The 
maximum negative energy occurs when $r \approx 2$, where $R-n \approx 0.3$.} 
\label{fig:sv_v} 
\end{figure}

A limit of special interest is when $\alpha =0$ and we have the superposition 
of the vacuum with a squeezed vacuum state. In this case,
\begin{equation}
N^2 = 
\left[ 1+|\eta|^2 +2 {\rm Re}(\eta) \, \sqrt{{\rm sech}\, r}\,\right]^{-1}\,,
\end{equation}
\begin{equation}
n = \langle a^\dagger a \rangle = N^2\, \sinh^2 r \, ,
\end{equation}
and
\begin{equation}
\langle a^2 \rangle = -N^2\,\sinh r \cosh r \, {\rm e}^{i \delta}\, 
\left[1+ \eta^*\, ({\rm sech}\, r)^\frac{5}{2} \right]\,.
\end{equation}
The $\alpha = 0$ curve in Fig.~\ref{fig:scs} is this limit for $\eta =1$.
If $\eta$ is real and negative, $\eta =-|\eta|$, we then have
\begin{equation}
R = |\langle a^2 \rangle| = N^2 \sinh r \cosh r \,
\left| 1- |\eta|\, ({\rm sech}\, r)^\frac{5}{2} \right| \, ,
\end{equation}
and
\begin{equation}
R -n = \frac{\sinh r\,\left[\cosh r \left| 1 -
|\eta|\,({\rm sech}\, r)^\frac{5}{2} \right| -\sinh r \right]}
{1+|\eta|^2 -2 |\eta| \, \sqrt{{\rm sech}\, r}} \,. \label{eq:R-n_svv}
\end{equation}
In the case that $\eta =-1$ the right-hand side of Eq.~(\ref{eq:R-n_svv}) 
is plotted as a 
function of $r$ in Fig.~\ref{fig:sv_v}. 
Here we find the maximum negative energy,
$R-n \approx 0.3$ at $r \approx 2$. This is slightly less negative energy
than can be found in a single squeezed vacuum state. Note that as $r 
\rightarrow 0$, this state becomes $|2 \rangle$, a two-particle state with 
positive energy density everywhere. This is the reason that the behavior
in  Fig.~\ref{fig:sv_v} differs from the $\alpha =0$ curve in 
Fig.~\ref{fig:scs}. In the latter case, $\eta = 1$, and there is negative
energy for all values of $r$.

\section{Two-Mode Entangled States}
\label{sec:entangled}

In this section, we will consider several examples of entangled states
involving two modes.

\subsection{An Entangled Squeezed State - the Barnett-Radmore State}
\label{sec:BR}

Our first example of a two-mode entangled squeezed state was described by 
Barnett and Radmore~\cite{BR} and is defined by 
\begin{equation}
|\psi\rangle = S_{AB}\,|0\rangle\,, 
\end{equation}
where $|0\rangle$ is the vacuum state for both modes, and
\begin{equation}
S_{AB} = {\rm e}^{(\xi^* a b - \xi a^\dagger b^\dagger)}
\end{equation}
is a two-mode squeeze operator. If one were to expand the state $|\psi\rangle$
in terms of number eigenstates, the expansion would contain states with an 
even total number of particles, with half of these particles in each mode.
One has the following identities~\cite{BR}
\begin{eqnarray}
S_{AB}(-\xi)\, a S_{AB}(\xi) = 
a \cosh r -b^\dagger \,{\rm e}^{i\delta}\,\sinh r
                                               \nonumber \\
S_{AB}(-\xi)\, a^\dagger S_{AB}(\xi) = 
a^\dagger \cosh r - b \,{\rm e}^{-i\delta}\,\sinh r
                                               \nonumber \\
S_{AB}(-\xi)\, b S_{AB}(\xi) = b\cosh r -a^\dagger \,{\rm e}^{i\delta}\,\sinh r
                                               \nonumber \\
S_{AB}(-\xi)\,b^\dagger  S_{AB}(\xi) = 
b^\dagger\cosh r -a\,{\rm e}^{-i\delta}\,\sinh r \, .
\end{eqnarray}
Note that                                               
\begin{equation}
S_{AB}^\dagger(\xi) =S_{AB}^{-1}(\xi) =S_{AB}(-\xi)\,. 
\end{equation}
One may use these relations to show that
\begin{equation}
n_1 = n_2 = \sinh^2 r\,, \quad R_1=R_2=R_3=0\,, \quad
R_4 = \sinh r\, \cosh r \,,\; {\rm and} \; \gamma_4 =\delta +\pi\,.
\end{equation}
The minimum energy density in this state is
\begin{equation}
\rho_{\rm min}(BR) = -\frac{\sinh r}{V} \; [ 2 \sqrt{\omega_1 \omega_2}\,\cosh r
-(\omega_1 +\omega_2) \sinh r ]\,.
\end{equation}
This is never more negative than the minimum energy density that would be
obtained if the two modes were individually in squeezed vacuum states. 
The latter
energy density is $\rho_{\rm min}(2SQ) = -(\omega_1+\omega_2)(R-n)/V$, where 
$R-n$ is given by Eq.~(\ref{eq:Rnsqvac}). Thus we can write
 \begin{equation}
\rho_{\rm min}(BR) - \rho_{\rm min}(2SQ) = \frac{\sinh r\, \cosh r}{V} \;
(\sqrt{\omega_1} - \sqrt{\omega_2})^2 \,,
\end{equation}
which is always non-negative and approached zero only when the two modes 
have nearly the same frequency.

\subsection{An Second Entangled Squeezed State - the Zhang State}
\label{sec:Zhang}

In this subsection, we will consider a second possibility for an entangled
two-mode squeezed state, which was discussed by Zhang~\cite{Zhang}.
This state is defined by
\begin{equation}
|\psi\rangle = N\left(|\bar{\xi}\rangle_a \, |\bar{\eta}\rangle_b +
{\rm e}^{i\theta}\, |{\xi}\rangle_a \, |{\eta}\rangle_b \right) \,,
\label{eq:Zhang}
\end{equation}
where $|\bar{\xi}\rangle_a$ and $|{\xi}\rangle_a$ are single-mode squeezed
vacuum states for mode $a$, and  $|\bar{\eta}\rangle_b$ and $|{\eta}\rangle_b$
 are such states for mode $b$. In general, $\xi$, $\bar{\xi}$,  $\eta$, 
and $\bar{\eta}$ can be four arbitrary complex parameters. However, we will
restrict our attention to the case where they are real and satisfy
\begin{equation}
\xi = \eta = - \bar{\xi} = - \bar{\eta} \,.  \label{eq:Zhang1} 
\end{equation}
In this case,
\begin{equation}
N = \left\{ 2 [1+ {\rm Re} ({\rm e}^{i\theta}\, 
 \langle -\xi |{\xi}\rangle_a \, \langle -
\eta |{\eta}\rangle_b)]\right\}^{-\frac{1}{2}}  
=\left[ 2\,(1+\cos \theta \,{\rm sech}\, 2r) \right]^{-\frac{1}{2}} \,,
\label{eq:ZhangN}
\end{equation}
where we have used Eq.~(\ref{eq:sech}) for each of the two modes.
Similarly, we find
\begin{equation}
n_1=n_2 = 2\,N^2\,\sinh^2 r \,
\left[1 - \frac{\cos \theta}{(\cosh 2r)^\frac{3}{2}} \right] \,,
\label{eq:Zhangn}
\end{equation}
and 
\begin{equation}
R_1=R_2 = N^2\, |\sin \theta|\; \frac{\tanh 2r}{\sqrt{\cosh 2r}} \,.
\label{eq:ZhangR}
\end{equation}
Here we have use Eqs.~(\ref{eq:ada}) and (\ref{eq:a2}), as well as the 
identity $\sinh^2 r + \cosh^2 r = \cosh (2r)$. 
In addition, we find $R_3=R_4 =0$ and $\gamma_1= \gamma_2= -\pi/2$.

In this case, the energy density, Eq.~(\ref{eq:2mode_rho}), becomes 
\begin{equation}
\rho = \frac{1}{V}\,\biggl( n_1\, (\omega_1 +  \omega_2)
+ R_1\, \left\{\omega_1\, \cos[2( \mathbf{k}_1 \cdot \mathbf{x} - 
\omega_1 t)+\gamma_1]
+  \omega_2\, \cos[2 (\mathbf{k}_2 \cdot \mathbf{x} - \omega_2 t)+\gamma_1]
  \right\} \biggr)
\end{equation} 
We can always choose the spatial position $\mathbf{x}$ and time $t$ so
as to make both cosine functions equal to $-1$, in which case we achieve
the minimum allowed energy density in this state of
\begin{equation}
\rho_{min} = - \frac{\omega_1 +  \omega_2}{V}\; (R_1 -n_1)\,.
\end{equation}

\begin{figure} 
\begin{center} 
\leavevmode\epsfysize=8cm\epsffile{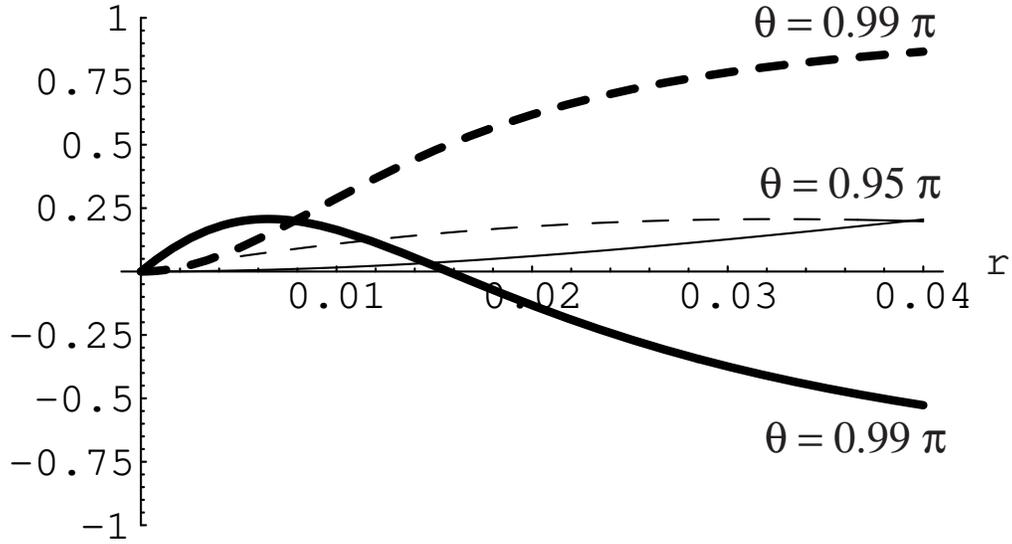} 
\end{center} 
\caption{ The quantities $R_1 -n_1$ (solid lines) and $n_1$ (dashed lines) 
are plotted for two values of $\theta$ as  
functions of $r$ for the entangled squeezed state defined in 
Eqs.~(\ref{eq:Zhang}) 
and (\ref{eq:Zhang1}). In the limit that $\theta$ is close to $\pi$ , one
can have appreciable negative energy at small values of $r$. 
We see that the peak
negative energy, $R_1-n_1 \approx 0.25$ occurs at $r \approx 0.007$ for
$\theta = 0.99 \pi$ and at $r \approx 0.035$ for $\theta = 0.95 \pi$, 
whereas the
mean particle number is about the same for both cases, $n_1 \approx 0.2$.  } 
\label{fig:Zhang} 
\end{figure} 

From Eqs.~(\ref{eq:ZhangN}), (\ref{eq:Zhangn}) and 
(\ref{eq:ZhangR}), we find $R_1 -n_1$ as a function of $\theta$ and
$r$. In general, the behavior of this entangled state is 
similar to that of the superposed squeezed vacua illustrated in 
Fig.~\ref{fig:2sq}. However, there is one limit of particular interest,
which is when $r \ll 1$ and $0 < |\pi - \theta| \ll 1$. (Note that if 
$\theta =\pi$, then $R_1 =0$, and there is no negative energy.)
If we take the
limit $r \ll 1$, for fixed $\theta \not= \pi$, then we find the asymptotic 
forms
\begin{equation}
n_1 \sim \frac{1 -\cos \theta}{1+\cos \theta}\, r^2 \,,
\end{equation}
and
\begin{equation}
R_1-n_1 \sim  \frac{|\sin \theta|}{1+\cos \theta}\, r \,.
\end{equation}
In the case that $0 < |\pi - \theta| \ll 1$, the coefficient in the expression
for $n_1$ can be large, so we can have an unusually large particle number
in relation to the value of $r$. The quantities $R_1-n_1$ and $n_1$ are
plotted in Fig.~\ref{fig:Zhang} for two values of $\theta$ close to $\pi$.
In this case, we can have a reasonable amount of negative energy at very
small values of the squeeze parameter, $r$.

\subsection{Entangled Coherent States}
\label{sec:en_coherent}

\begin{figure} 
\begin{center} 
\leavevmode\epsfysize=8cm\epsffile{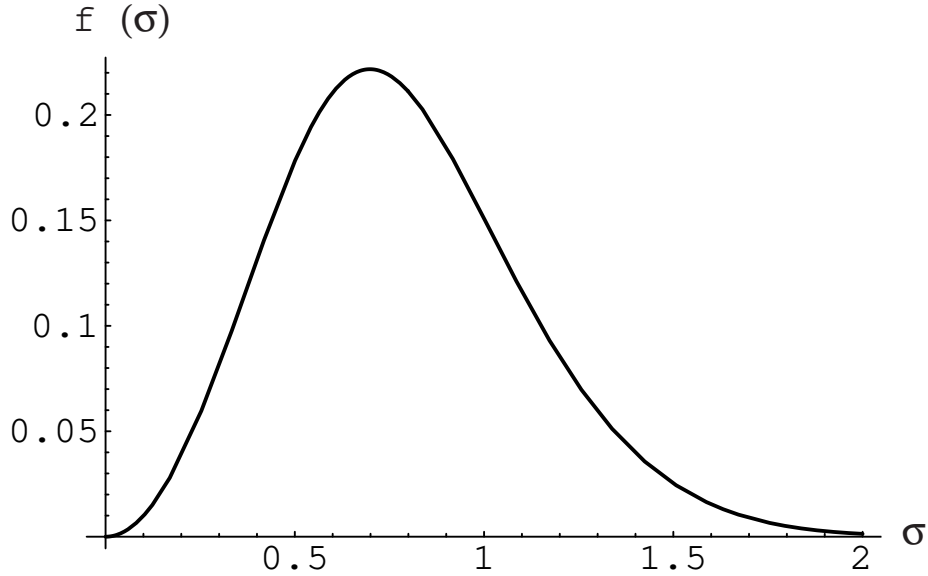} 
\end{center} 
\caption{The function $f(\sigma)$, given by Eq.~(\ref{eq:fsig}), is plotted. 
Its maximum,
at $\sigma \approx 0.7$, describes the case of maximal negative energy 
density for the entangled coherent state defined in 
Eqs.~(\ref{eq:en_coherent}), (\ref{eq:en_coherent1}) and 
(\ref{eq:en_coherent2}).   } 
\label{fig:en_coherent} 
\end{figure}

In this subsection, we consider a state of the same form as that in
Eq.~(\ref{eq:Zhang}), but involving entangled coherent states for two modes,
which was also discussed by Zhang~\cite{Zhang}.  Let
\begin{equation}
|\psi\rangle = N\left(|\alpha\rangle_a \, |\beta\rangle_b +
{\rm e}^{i\theta}\, |{\alpha'}\rangle_a \, |\beta' \rangle_b \right) \,,
\label{eq:en_coherent}
\end{equation}
where $|\alpha\rangle_a$, {\it etc} are single-mode coherent states. 
We will restrict our attention to the case where the magnitudes of the four
complex coherent state parameters are all equal, and $\alpha'=-\alpha$
and $\beta' =-\beta$. Thus
\begin{equation}
|\alpha|=|\beta|=|\alpha'|=|\beta'|= \sigma \, , \label{eq:en_coherent1}
\end{equation}
and
\begin{equation}
\delta_1 -\delta_1' = \pm \pi \, , \quad \delta_2 -\delta_2'= \pm \pi \,.
\label{eq:en_coherent2}
\end{equation}
Here $\delta_1,  \delta_1', \delta_2,  \delta_2'$ are the phases of
$\alpha, \alpha', \beta, \beta'$, respectively. In this case, we find
\begin{equation}
N = \left[2(1+\cos \theta\, {\rm e}^{-4 \sigma^2})\right]^{-\frac{1}{2}} \,,
\end{equation}
and 
\begin{eqnarray}
n_1 &=& n_2 = 2 \sigma^2\, N^2\, (1 - \cos \theta \, {\rm e}^{-2 \sigma^2})\,,
                                     \nonumber \\
R_1 &=& R_2 = 2 \sigma^2\, N^2\, (1 + \cos \theta \, {\rm e}^{-2 \sigma^2})\,,
                                     \nonumber \\
R_3 &=&  \sigma^2\, N^2\, (1 - \cos \theta \, {\rm e}^{-4 \sigma^2})\,,
                                     \nonumber \\
R_4 &=& \sigma^2 \,,
\end{eqnarray}
as well as $\gamma_1 = 2\delta_1'$, $\gamma_2 = 2\delta_2'$,
$\gamma_3 = \delta_2 - \delta_1$, and $\gamma_4 = \delta_1 + \delta_2$.
Let $\phi_1 =      \bf{k}_1 \cdot     \bf{x} - \omega_1 t$ and
 $\phi_2 =      \bf{k}_2 \cdot     \bf{x} - \omega_2 t$. We then set
\begin{equation}
\phi_1 +\delta_1 = \phi_2 +\delta_2 = \frac{\pi}{2} \, ,
\end{equation}
which can always be done by a suitable choice of $\mathbf{x}$ and $t$.
The energy density for a two-mode traveling wave state, 
Eq.~(\ref{eq:2mode_rho}) now becomes
\begin{equation}
\rho = \frac{1}{V}\, \left[n_1 \omega_1 + n_2 \omega_2 - R_1 \omega_1
- R_2 \omega_2 +(R_3-R_4) \,  \sqrt{\omega_1 \omega_2}\, 
(1+ \mathbf{\hat{k}}_1 \cdot \mathbf{\hat{k}}_2)\, \right]\,.
\end{equation}
If the two modes are close in wavenumber, so that $\omega_1 \approx 
\omega_2 = \omega$ and     $\bf{\hat{k}}_1 \approx     \bf{\hat{k}}_2$,
and we set $\theta =0$ then 
\begin{equation}
\rho = - \frac{4\, \omega}{V}\, f(\sigma) \,,
\end{equation}
where
\begin{equation}
f(\sigma) = \frac{\sigma^2\, {\rm e}^{-2 \sigma^2}\, 
(1 +  {\rm e}^{-2 \sigma^2})}{ 1 +  {\rm e}^{-4 \sigma^2} }\,. \label{eq:fsig}
\end{equation}
The function $f(\sigma)$ is plotted in Fig.~\ref{fig:en_coherent}, 
where we see that it attains
a maximum value of about $0.22$ at $\sigma \approx 0.7$. This corresponds
to a negative energy density of $\rho \approx -0.88 \omega/V$, which is about
three times as negative as the maximally negative energy density found in
the superposed coherent states discussed in Sect.~\ref{sec:SCS}.

\section{Summary}
\label{sec:Sum}

In this paper we have developed a formalism for parameterizing the energy
density in states of a massless scalar field in which either one or two modes
 are excited. We found explicit expressions for the energy density for the 
cases of traveling waves and of standing waves in one spatial direction.
In all cases, the maximum negative energy density which can be achieved in 
a given state can be expressed in terms of our parameters.

We next applied this approach to find the maximum negative energy density
in several states, including some states which are of current interest
in quantum optics. For the case of a single mode, we considered three
possible superposition states: (1) two coherent states, (2) two squeezed
vacuum states, and (3) a coherent state and a squeezed vacuum state.  
The superposition of two coherent states can be described as a Schr{\"o}dinger
 ``cat state'' in the sense that it would be a superposition of two classical
configurations in the limit of large coherent state parameter. Here we find
that the maximal negative energy density is achieved with mean photon numbers
slightly less than one. This is an example where a quantum superposition state
has negative energy density, even though each component of the superposition
would have positive energy density by itself. In the case of the superposition
of two squeezed vacuum states, one finds the opposite effect. Although here 
the superposition state does has negative energy density, it is somewhat less
negative than in the case of a single squeezed vacuum state. Furthermore,
the most negative energy density now occurs for small mean photon number, as
opposed to large number in the case of a single squeezed vacuum state. 
In the case of a superposition of a coherent state and a squeezed vacuum 
state, we find that for fixed coherent state parameter, there is negative
energy density for small squeeze parameter, and again for larger values, but
there is an intermediate range where the energy density is always positive.

We next examined some two-mode states involving entanglement between the two
modes, including two examples of entangled squeezed vacuum states. The
first example, the Barnett-Radmore state~\cite{BR}, exhibits somewhat less
negative
energy density than would be found if each mode were separately in a squeezed
vacuum state. In the second example, the Zhang state~\cite{Zhang},  we
find results similar to those in the superposition of squeezed vacuum 
states. There is negative energy in the Zhang state, but only for small
mean particle numbers. Finally, we examined a two-mode entangled coherent 
state, which also exhibits  negative energy for  small mean particle number.
It is also similar to the case of a superposition of coherent states of
a single mode, but the entangled state has somewhat more  negative
energy density. 

One of the motivations for this investigation is to draw links between
theoretical studies of violations of the weak energy condition, and 
experimental work in quantum optics. We hope that this line of work
will lead to further experimental studies of subvacuum phenomena.

\begin{acknowledgments}
We would like to thank Piotr Marecki for useful discussions.
 This work was supported in part by the National Science Foundation 
under Grant PHY-0555754 to LHF.
\end{acknowledgments}

\appendix
\section{}

In this appendix, we will calculate some of the matrix elements of operators
such as $a^\dagger\,a$ and $a^2$ which are needed to find the energy density
in the states treated in this paper. We begin with matrix elements between
squeezed vacuum states. The diagonal matrix elements 
$\langle \xi |a^\dagger\,a |\xi \rangle$ and $\langle \xi |a^2 |\xi \rangle$ 
are given by Eq.~(\ref{eq:diag_sq}). We need off-diagonal matrix elements
of the form $\langle- \xi |a^\dagger\,a |\xi \rangle$ and 
$\langle -\xi |a^2 |\xi \rangle$, where $\xi$ is real. If we set $\delta=0$, so
that $\xi =r$, then Eq.~(\ref{eq:xi}) becomes  
\begin{equation}
|\xi \rangle = 
\sqrt{{\rm sech}r} \, \sum_{n=0}^\infty \, \frac{\sqrt{(2n)!}}{n!} 
{\biggl(-\frac{1}{2} {\rm tanh}r \biggr)}^n \,|2n \rangle \,.
\label{eq:xi2}
\end{equation}
This leads to the result 
\begin{equation}
\langle -\xi | a^\dagger a  |\xi \rangle = 
\langle \xi | a^\dagger a  |-\xi \rangle =
2 \,{\rm sech}r \, \sum_{n=1}^\infty \, \frac{(2n)!}{n!(n-1)!} \, {(-1)}^n 
{\biggl(\frac{1}{2} {\rm tanh}r \biggr)}^{2n} \,.
\label{eq:SadagaS}
\end{equation}  
Use the fact that 
\begin{equation}
\sum_{n=1}^\infty \, 
\frac{(2n)!}{n!(n-1)!}\, {(-1)}^n \,
{\biggl(\frac{1}{2} x \biggr)}^{(2n-2)} = 
-\frac{2}{{(1+x^2)}^{3/2}} \,,
\label{eq:xsum}
\end{equation}
to find   
\begin{equation}
\langle -\xi | a^\dagger a  |\xi \rangle = 
\langle \xi | a^\dagger a  |-\xi \rangle =
-\frac{{\rm sech}r \,{{\rm tanh}^2r}}{{(1+{\rm tanh}^2 r)}^{3/2}} \,.
                                         \label{eq:ada}
\end{equation}  
Similarly, we may use  Eq.~(\ref{eq:xi2}) to show that
\begin{equation}
\langle -\xi | a^2 |\xi \rangle = -\langle \xi | a^2 |-\xi \rangle =
{\rm sech}r \, \sum_{n=1}^\infty \, 
\frac{(2n)!}{n!(n-1)!}\, {(-1)}^n \,
{\biggl(\frac{1}{2} {\rm tanh}r \biggr)}^{(2n-1)}
= -\frac{{\rm sech}r \,{{\rm tanh}r}}{{(1+{\rm tanh} r)}^{3/2}} \,.
                                          \label{eq:a2}
\end{equation} 
Because these matrix elements are real, we have that
\begin{equation}
\langle -\xi | (a^\dagger)^2 |\xi \rangle = 
\langle \xi | (a^\dagger)^2  |-\xi \rangle = 
\langle -\xi | a^2 |\xi \rangle \,.
\end{equation}
In the present case, the squeeze operator is  
\begin{equation}
S(\xi) = S(r) = {\rm e}^{\frac{1}{2}r [a^2 - {(a^\dagger)}^2]}\,.
\label{eq:Squeezer}
\end{equation}
From this relation, we see that
\begin{equation}
\langle -\xi |\xi \rangle = \langle \xi |-\xi \rangle = 
\langle 0 |S^2(r)| 0 \rangle = 
\langle 0 |S(2r)| 0 \rangle = \sqrt{{\rm sech}(2r)}\,.
\label{eq:sech}
\end{equation}

Next we derive the matrix elements involving both a coherent state and a
squeezed vacuum state that are needed in Sect.~\ref{sec:CSV}. A coherent
state may be represented in terms of number eigenstates as~\cite{BR}
\begin{equation}
|\alpha \rangle = {\rm e}^{-\frac{1}{2} |\alpha|^2}\;
\sum_{\ell=0}^\infty \frac{\alpha^\ell}{\sqrt{\ell !}}\, |\ell \rangle \,.
\end{equation}
This may be combined with Eq.~(\ref{eq:xi}) to show that
\begin{eqnarray}
\langle \xi|\alpha \rangle = \langle \alpha | \xi \rangle^* &=&
{\rm e}^{-\frac{1}{2} |\alpha|^2} \, \sqrt{{\rm sech}r}\,
\sum_{n=0}^\infty \frac{\alpha^{2n}}{{n !}}\,
[-\frac{1}{2} {\rm e}^{-i\delta} \tanh r]^n \nonumber \\
&=& {\rm e}^{-\frac{1}{2} |\alpha|^2} \, \sqrt{{\rm sech}r}\,
\exp\left(-\frac{1}{2} {\rm e}^{-i\delta} \alpha^2  \tanh r \right)\,,
                                            \label{eq:in_prod}
\end{eqnarray}
and
\begin{equation}
\langle \xi |a^\dagger a |\alpha \rangle = 
\langle \alpha | a^\dagger a  |\xi \rangle^* = 
-{\rm e}^{-\frac{1}{2} |\alpha|^2} \, \sqrt{{\rm sech}r}\,
{\rm e}^{-i\delta} \alpha^2  \tanh r \,
\exp\left(-\frac{1}{2} {\rm e}^{-i\delta} \alpha^2  \tanh r \right)\,.
                                             \label{eq:ada2}
\end{equation}
Note that
\begin{equation}
\langle \xi | a^2 |\alpha \rangle = 
\langle \alpha | (a^\dagger)^2  |\xi \rangle^* = 
\alpha^2\, \langle \xi|\alpha \rangle \,.
\end{equation}
Finally, we show that
\begin{eqnarray}
\langle \xi | (a^\dagger)^2  |\alpha \rangle &=& 
\langle \alpha | a^2  |\xi \rangle^* = 
{\rm e}^{-\frac{1}{2} |\alpha|^2} \, \sqrt{{\rm sech}r}\,
\sum_{n=1}^\infty \frac{2n(2n-1)}{n!}\, \alpha^{2n-2} \,  
\left(-\frac{1}{2} {\rm e}^{-i\delta}  \tanh r \right)^n \nonumber \\
&=& {\rm e}^{-\frac{1}{2} |\alpha|^2} \, \sqrt{{\rm sech}r}\,
\frac{\rm d^2}{{\rm d} \alpha^2} \;
\sum_{n=0}^\infty \frac{\alpha^{2n}}{n!}\,  
\left(-\frac{1}{2} {\rm e}^{-i\delta}  \tanh r \right)^n \nonumber \\
&=&  {\rm e}^{-\frac{1}{2} |\alpha|^2} \, \sqrt{{\rm sech}r}\,
\frac{\rm d^2}{{\rm d} \alpha^2} \;
\exp\left(-\frac{1}{2} {\rm e}^{-i\delta} \alpha^2  \tanh r \right)
\nonumber \\
&=& {\rm e}^{-\frac{1}{2} |\alpha|^2}\,  \sqrt{{\rm sech}r} \,
\left({\rm e}^{-i\delta} \alpha^2  \tanh r -1 \right) \, 
{\rm e}^{-i\delta}\, \tanh r  \nonumber \\  
&\times& \exp\left(-\frac{1}{2} {\rm e}^{-i\delta} \alpha^2  \tanh r \right) .
                                           \label{eq:ad2}
\end{eqnarray}

\end{document}